# All-optical detection of periodic structure of chalcogenide superlattice using coherent folded acoustic phonons

*Takara Suzuki, Yuta Saito, Paul Fons, Alexander V. Kolobov, Junji Tominaga and Muneaki Hase[1*]*

T. Suzuki, M. Hase
Division of Applied Physics, Faculty of Pure and Applied Sciences, University of Tsukuba, 1-1-1 Tennodai, Tsukuba 305-8573, Japan.
E-mail: mhase@bk.tsukuba.ac.jp

Y. Saito, P. Fons, A. V. Kolobov, J. Tominaga
Nanoelectronics Research Institute, National Institute of Advanced Industrial Science and Technology, Tsukuba Central 5, 1-1-1 Higashi, Tsukuba 305-8565, Japan.

**ABSTRACT**

**Chalcogenide superlattices (SL) consist of alternate stacking of GeTe and $Sb_2Te_3$ layers. The structure can become a 3D topological insulator depending on the constituent layer thicknesses, making the design of the SL period a central issue for advancing chalcogenide SL as potential candidates for spin devices as well as for optimization of the current generation of phase-change memory devices. Here we explore the periodic structure of chalcogenide SL by observing coherent folded longitudinal acoustic (FLA) phonons excited by femtosecond laser pulse irradiation. The peak frequency of the FLA modes was observed to change upon variation of the thickness of the GeTe layer, which was well reproduced by means of an elastic continuum model. In addition, a new SL structure is unveiled for a sample consisting of thin GeTe and $Sb_2Te_3$ layers, which suggests intermixing of Ge atoms. This all-optical technique based on observation of coherent FLA modes offers a non-destructive characterization of superlattice structures at atomic resolution.**

# 1. Introduction

The class of chalcogenide glass materials that shows significant changes in optical and electrical properties upon an amorphous-to-crystalline phase transition has led to the development of large data capacities in modern optical data storage [1–3]. The transition between the two phases can be induced by irradiation with focused nanosecond laser pulses, leading to melt-quenching (amorphization) and annealing (crystallization), respectively. The use of this phenomena has led to the development of devices whose reliability allows more than $10^6$ write-erase cycles. Moreover chalcogenide glass materials have been applied to a new class of fiber-core materials [4], which show extreme optical properties in the mid-infrared wavelength range, a range critical for optical communications [5]. Recently, all-optical multilevel memory operation has been realized using chalcogenide glass materials over optical communication wavelengths, in which a write/erase pulse induces a phase-change leading to a change in transmission of a subsequent read pulse [6]. Thus chalcogenide glass phase-change materials (PCMs) have the potential for a wide range of industrial applications as optical materials.

Among the chalcogenide PCMs, $Ge_2Sb_2Te_5$ (GST225) is one of the best-performing Ge-Sb-Te alloys, and has already been commercialized as a recording layer for optical disks, such as DVD-RAM (digital versatile disk – random access memory). It is also anticipated that these materials will be utilized as a key elements of non-Von Neumann or neuromorphic computing, where the PCM serves as computational memory [7–9]. To reduce the switching energy in nonvolatile memories such as phase-change random access memory (PCRAM), Chong *et al.* proposed an SL-like PCRAM structure in which they considered the GST225 alloy system as a composite of pseudo-binary alloys, namely GeTe and $Sb_2Te_3$, with individual layers thick enough to maintain the characteristics of each composition [10]. Both faster switching times (< 5 ns) and lower programming currents (≈1/3) were found for the SL-like PCRAM. More recently, Simpson and Tominaga *et al.* introduced interfacial phase change memory (iPCM), consisting of the alternate stacking of GeTe and $Sb_2Te_3$ layers each a few unit cells thick (**Figure 1**a) [11]. iPCM shows better energy efficiency ($\approx$1/9 of conventional PCRAM), faster phase change kinetics, and outstanding cyclability ($>10^9$) compared to GST225 alloy-based PCRAM devices. It is believed that by applying electrical or optical pulses, crystal–crystal transitions between the low-resistivity (SET) and the high-resistivity (RESET) phases without an intermediate melting state are possible in chalcogenide SL [11,12]. The switching mechanism proposed is different from that of conventional GST225, due to the presence of a ferroelectric atomic sequences in the GeTe layer [13–15]; recently strain-engineered diffusive atomic switching in chalcogenide SL has also been proposed [15]. Note that the SL structure consists of a ferroelectric normal insulator, GeTe, and a topological insulator (TI), $Sb_2Te_3$, and for certain structures it was shown using *ab initio* calculations that the RESET phase becomes a Dirac semimetal, while the SET phase becomes a Weyl semimetal [16]. As can be anticipated from the characteristics of a TI, the band gap closes to give rise to a Dirac cone for certain thickness combinations or applied stress of the GeTe and $Sb_2Te_3$ layers [17,18].

However, it has recently been argued that the periodicity of the SL structure may significantly differ from the proposed stacking of GeTe and $Sb_2Te_3$ [19–22]. Such variations would give rise to aperiodic van der Waals (vdW) gaps in the SL structure, which may lead to degradation of device performance. Variations in vdW layers periodicity were directly observed by high angle annular dark field (HAADF) scanning transmission electron microscopy (STEM) suggesting the inter-diffusion of Ge and Sb atoms results in the formation of ternary vdW layers [20,22–24]. These microscope techniques are sometimes useful to evaluate the formation of aperiodic vdW gaps, however, such measurements are destructive due to the need to produce thin cross sections. Therefore, alternative non-destructive methods are urgently needed to evaluate SL structures without mechanical contact.

In this work, we demonstrate an all-optical investigation of SL periodicity in an SL structure through observation of coherent folded longitudinal acoustic (FLA) phonons using coherent phonon spectroscopy (CPS). To date, these modes have not been examined in chalcogenide SL. We observed variations in the FLA peak and its corresponding frequency shift with the thickness of the GeTe layers, allowing the determination of the SL structure of the samples. In addition, by comparing our results with Rytov's elastic continuum model, we found nearly perfect periodicity of the vdW gaps. Thus, observation of the FLA mode is a powerful tool for evaluating the periodicity of a chalcogenide SL [25-27].

In an SL structure with a period $D$ the phonon dispersion forms a mini-Brillouin zone in reciprocal space, since $D$ is longer than the lattice constant $a$ [28,29]. The zone edge in reciprocal space is defined by $k = \pi/D$, where $k$ is the wavevector, leading to a zone-folding effect. As a result of the zone-folding, the acoustic phonon dispersion curve is folded at $k = \pi/D$ and exhibits a zigzag line shape (Figure 1b), making it possible to observe the FLA phonon at $k \approx 0$ by Raman scattering and CPS [30–32]. Thus, we expect to observe coherent FLA modes at $k_{laser} = 4n\pi/\lambda$ (where $n$ and $\lambda$ are the refractive index of the sample and the probe wavelength, respectively), if the samples possess an SL structure.

## 2. Experimental Section

**2.1** *Sample Fabrication*: The chalcogenide SL samples, (GeTe)$_{n=2, 4, 6}$/Sb$_2$Te$_3$, were fabricated at 230 ℃ using radio-frequency magnetron sputtering and were deposited on Si (100) substrates, where n=2, 4, and 6 correspond to the thickness of GeTe being 0.8, 1.6, and 2.4 nm, respectively, herein referred to as samples A, B, and C. The Sb$_2$Te$_3$ layer thickness was fixed at 1 nm for all samples. Before the growth of the superlattice, the Si substrates were sputter-cleaned with Ar plasma and a 3-nm-thick Sb$_2$Te$_3$ layer was deposited at room temperature and subsequently annealed at the growth temperature to form a seed layer [33,34]. The ZnS-SiO$_2$ (20 nm) layer was grown in the same sputtering chamber to prevent oxidation.

**2.2** *Coherent phonon spectroscopy*: We utilized a pump-probe technique to obtain the coherent phonon spectra by the Fourier transformation (FT) of the transient reflectivity change. Femtosecond laser pulses from a mode locked Ti:sapphire laser (pulse width 20 fs, central wavelength 820 nm, repetition rate 80 MHz, average power 320 mW) were used as a light source. The irradiated pump pulses impulsively generated coherent phonons, which results in the reflectivity oscillations observed by the delayed probe pulse. To avoid pump and probe interference, the polarizations of the pulses were horizontal and vertical, respectively. The photo induced reflectivity change ($\Delta R/R$) was recorded as a function of the time delay between the two pulses, which was scanned by a horizontally shaking mirror at a 19.5 Hz scan frequency [35,36]. Experiments were conducted at room temperature with a constant pump power (140 mW) and spot size (~70 μm), from which the pump fluence was 0.14 mJ/cm$^2$ over the experiments. All samples measured in this study were as-deposited films without post annealing.

**2.3** *First principle calculation of phonon dispersion*: First principle calculations of the phonon dispersions were carried out using the plane-wave code VASP [37] in conjunction with the PHONOPY package [38]. Calculations were carried out at 0K using the generalized gradient approximation (GGA) of Perdew et al [39] and projector augmented wave pseudopotentials [39,40]. The Ge, Te, and Sb projector augmented wave (PAW) pseudopotentials explicitly included $4s^24p^2$, $5s^25p^4$, and $5s^25p^3$ as valence electrons, respectively. A plane-wave cutoff of 500 eV was used throughout with a 3×3×2 Monkhorst-Pack grid. After relaxation, the force constants of the dynamical equation for both the GeTe and Sb$_2$Te$_3$ structures in the hexagonal setting were calculated using density functional perturbation theory using 2×2×1 supercells to take into account longer range interactions. The simultaneous convergence criteria were set to the following values: an energy of 1×10$^{-8}$ eV/atom and maximum force and stress tensor components of 0.01 eV/Å. The sound velocity was determined by fitting the slope of the acoustic branch in the vicinity of the Γ point.

## 3. Results and Discussion

To examine the effect of the SL period ($D$) on the coherent FLA modes, we prepared three chalcogenide SL samples, named A, B, and C, in which the GeTe layer thickness ($d$) was varied from 0.8 to 2.4 nm, and the thickness of the $Sb_2Te_3$ layer was fixed to 1.0 nm for all samples. The transient reflectivity change ($\Delta R/R$) signal for sample A ($d_{GeTe}$ = 0.8 nm) is presented in **Figure 2**a. The time-domain data was fit with a double-exponential decaying function to subtract the electronic response. In the Fourier transformed (FT) spectra of the residual shown in Figure 2b, three distinct peaks are clearly observed. The two peaks at higher frequency arise from the $A_1$ (3.16 THz) mode confined in the GeTe layer and the $A^2_{1g}$ (5.32 THz) mode confined in the $Sb_2Te_3$ layer, both of which were detected in previous experiments [41,42]. On the other hand, the lower frequency peak is identified as an FLA peak (1.62 THz). In the following, we confirmed that the lower frequency peak is an FLA mode by comparison of the spectra of the three samples shown in **Figure 3**. Most importantly, the significant peak shift of the FLA mode from 1.62 to 0.99 THz demonstrates that the zone-folding effect, which depends on $D$ and is a signature of the FLA, is clearly observed. The FLA peaks did not show a doublet or triplet peak structure, a structure which has been previously observed in Raman and CPS measurements in III-V semiconducting SL [28,32]. However, we note that a possible reason for the singlet peak nature could be small inhomogeneites in the vdW gap period.

Moreover, in the reflection-mode pump-probe optical layout used here, a small contribution from forward scattering can be seen due to interfacial reflection of the transmitted light [32,43,44]. Therefore, coherent FLA modes at $k$ = 0 may be observed in addition to those at $k_{laser}$. Note that an additional FLA peak can be seen at ≈0.7 THz for sample A (Figure 3), an observation which will be discussed in more detail in a later section.

To evaluate the SL structure, the experimental FLA frequency was compared with the theoretical value (**Figure 4**). On the basis of the elastic continuum model by Rytov et al. [29,45], we calculated the zone-folded LA phonon dispersion curve given by,

$$\cos(kD) = \cos\left[\omega\left(\frac{d_A}{v_A} + \frac{d_B}{v_B}\right)\right] - \frac{(\rho_B v_B - \rho_A v_A)^2}{2\rho_B v_B \rho_A v_A} \sin\left(\frac{\omega d_A}{v_A}\right) \sin\left(\frac{\omega d_B}{v_B}\right), \quad (1)$$

where $v$, $d$, $\omega$, and $k$ are the sound velocity, the layer thickness, the phonon frequency of each component, and the wave vector, respectively, whereas indices A and B correspond to the GeTe and $Sb_2Te_3$ layers, respectively. The sound velocities were obtained from bulk acoustic phonon dispersion curves at $k \approx 0$ calculated using first principles calculations. Since the frequency of the $A_1$ mode of the GeTe layer does not significantly change as the layer thickness increases (Fig. 3), we assume the effect of the strain on the velocity of sound in GeTe layer embedded in $Sb_2Te_3$ layers will be negligibly small. Contrary to a number of studies reporting the presence of aperiodic SL structure due to intermixing, good agreement was obtained for three chalcogenide SL samples in a comparison between the experimental and theoretical FLA mode frequencies calculated by Equation (1) (Figure 4). This unanticipated agreement indicates that SL structures fabricated by sputtering can realize the designed periodicity. A possible cause of the small deviations may arise from the calculation model and the actual SL structure grown.

The bulk LA phonon frequency in the Γ - Z direction of $Sb_2Te_3$ is much smaller than that of GeTe, and therefore, the observed FLA mode frequency, which should be between the two bulk curves, is slightly shifted from the zone-folded dispersion curves calculated from the elastic continuum model. In addition to this, since the bulk phonon dispersion curves become nonlinear in the large wavevector region, the elastic continuum model, which is based on linear phonon dispersion, cannot reproduce well the FLA phonon frequency when the period $D$ is small, i.e., $k = \pi/D$ is large [28]. Thus, the calculated FLA phonon frequency for sample

A, which has the smallest period among the three samples, deviates more from the experimental value compared to samples B and C.

Furthermore, the stacking of GeTe and $Sb_2Te_3$ will lead to a small difference in the FLA mode frequency. Recently, Kolobov et al. showed that an SL structure such as that of sample A, whose average composition corresponds to GST225, has an atomic sequence of Te-Sb-Te-Ge-Te-Ge-Te-Sb-Te in one vdW layer, namely the structure proposed by Kooi [46]. This is realized when about half of the deposited $Sb_2Te_3$ materials becomes capping layers for GeTe and the other half serves as a scaffold for the GeTe layers, namely, the GeTe layer lies between the two separated $Sb_2Te_3$ layers within one vdW layer. This causes some bond lengths to increase, resulting in a larger SL period. As mentioned above, a larger SL period results in a smaller first order FLA mode frequency. According to the literature [46], when GeTe layers are stacked on $Sb_2Te_3$ layers, Ge atoms face the $Sb_2Te_3$ layer and Te atoms act as a passivation layer, which results in a lower free energy compared to the case where Ge atoms form the terminating surface. In this structure, the Te-Te and Ge-Te distances become longer than those in bulk GST225 state. This may result in a small increase in the SL period.

Since the composition of sample A is the same as that of a GST225 alloy, we predict that expansion of the SL period is present in sample A. This is not the case for samples B and C due to the presence of a thicker GeTe layer and therefore they show good agreement between calculation and experiment.

In addition to the observation of the FLA modes, we obtained an unexpectedly large red shift in the $Sb_2Te_3$ $A^2_{1g}$ peak around 5 THz. As observed in Ref. [12] the $A^2_{1g}$ peak shows a small red shift as the thickness of the $Sb_2Te_3$ layers in the SL increase. These shifts may originate from strain induced by the increase in the thickness of the GeTe and $Sb_2Te_3$ layers; GeTe and $Sb_2Te_3$ layers experience tensile and compressive strain, respectively, due to their counterparts [15]. Therefore, we assume that increasing thickness results in a larger strain in each $Sb_2Te_3$ layer leading to reduced polycrystalline orientation.

Very recently, Cojocaru et al. have shown that the diffusion of Sb atoms into GeTe layers is much larger than that of Ge or Te atom into $Sb_2Te_3$ layers [47], supporting the results of Ref. [23]. Our results can be discussed based on these observations. As the GeTe layer becomes thicker, compressive strain in the $Sb_2Te_3$ layers increases [15]. On the other hand, in the GeTe layers, the tensile strain induced by the lattice mismatch is relatively small, and as the GeTe layers become thicker, the strain originating from the $Sb_2Te_3$ layers becomes even smaller. Therefore, we did not observe a significant peak shift in the GeTe $A_1$ mode whereas we observed a relatively large shift for the $Sb_2Te_3$ $A^2_{1g}$ mode. This observation corresponds to the fact that when Ge atoms diffuse into $Sb_2Te_3$ layers [21] the tensile strain [15] in the GeTe layer is relatively small compared to that in the $Sb_2Te_3$ layers. We note that the results for chalcogenide SL samples with varying $Sb_2Te_3$ thickness are not shown because the FLA peak is masked by the much stronger $Sb_2Te_3$ $A^1_{1g}$ mode [48] at ≈2THz.

Finally, we discuss a possible intermixed SL structure for the sample A, based on the observation of an additional FLA mode (≈0.7 THz) in Figure 3. The observation of the additional FLA peak cannot be explained by the SL structure model as designed, and we assume that there are other periodicities mixed in the $(GeTe)_2/Sb_2Te_3$ periodic structure. To explore the new periodicity, we calculated the FLA frequency for a $GeSb_2Te_4/Ge_3Sb_2Te_6$ SL (**Figure 5**a), resulting from the intermixing of Ge atoms. The diffusion stabilized $GeSb_2Te_4/Ge_3Sb_2Te_6$ SL structure was predicted by an optimization algorithm wrapped around DFT energy computations.[46,49] Based on the conservation of the number of atoms, it should be noted that one period of $GeSb_2Te_4/Ge_3Sb_2Te_6$ SL equals two periods of a $(GeTe)_2/Sb_2Te_3$ SL. Figure 5b shows very good agreement between the observed peak frequency and the calculation. Thus, our results strongly suggest the existence of a partial SL structure of $GeSb_2Te_4/Ge_3Sb_2Te_6$ due to the presence of atomic diffusion in the designed $(GeTe)_2/Sb_2Te_3$ chalcogenide SL.

As shown in Figure 3, unlike the case for sample A, we did not observe additional FLA modes for samples B and C. There is, however, also a possible intermixed structure candidate for the samples B and C. To study this possibility, we calculated the dispersion curves for $(GeTe)_{(0.4 \text{ or } 1.2 \text{ nm})}/(Ge_3Sb_2Te_6)_{(2.2 \text{ nm})}$ SL structures and the first-order FLA frequency is shown in **Figure S1** (supporting information). The calculated FLA frequencies based on these intermixed structures for samples B and C are significantly closer to those of the designed $(GeTe)_2/Sb_2Te_3$ structures. Therefore, it can be concluded that inter-diffusion of atoms occurred in samples B and C, and their FLA peaks contribute substantially to the broadening of the FLA peaks shown in Figure 4b, c.

## 4. Conclusion

In conclusion, we have observed coherent FLA modes in highly-oriented chalcogenide superlattice samples and found that the FLA frequency depends on the thickness of the GeTe layers and thereby the superlattice period. The current experimental results confirm the expected superlattice structure of the chalcogenide superlattices by means of an all-optical method. Furthermore, we experimentally uncovered a new periodic structure, $GeSb_2Te_4/Ge_3Sb_2Te_6$, when the thickness of the GeTe layer is ultrathin. This all-optical technique based on the coherent FLA modes offers a new pathway to characterize superlattice structures at atomic resolution, and will provide complementary information relevant to the currently most important issues, such as intermixing.

## Acknowledgements

This work was supported by CREST (NO. JPMJCR14F1), JST, Japan. We acknowledge Ms. R. Kondou for sample preparation.

**References**

[1] M. Wuttig, N. Yamada, *Nat. Mater.* **2007**, *6*, 824.

[2] S. R. Ovshinsky, *Phys. Rev. Lett.* **1968**, *21*, 1450.

[3] N. Yamada, E. Ohno, K. Nishiuchi, N. Akahira, M. Takao, *J. Appl. Phys.* **1991**, *69*, 2849.

[4] C. R. Petersen, U. Møller, I. Kubat, B. Zhou, S. Dupont, J. Ramsay, T. Benson, S. Sujecki, N. Abdel-Moneim, Z. Tang, D. Furniss, A. Seddon, O. Bang, *Nat. Photonics* **2014**, *8*, 830.

[5] C. Rios, P. Hosseini, C. D. Wright, H. Bhaskaran, *Adv. Mater* **2014**, *26*, 1372.

[6] C. Rios, M. Stegmaier, P. Hosseini, D. Wang, T. Scherer, C. D. Wright, H. Bhaskaran, W. H. P. Pernice, *Nat. Photonics* **2015**, *9*, 725.

[7] W. W. Koelmans, A. Sebastian, V. P. Jonnalagadda, D. Krebs, L. Dellmann, E. Eleftheriou, *Nat. Commun.* **2015**, *6*, 9183.

[8] J. Feldmann, M. Stegmaier, N. Gruhler, C. Riós, H. Bhaskaran, C. D. Wright, W. H. P. Pernice, *Nat. Commun.* **2017**, *8*, 1256.

[9] T. Tuma, A. Pantazi, M. Le Gallo, A. Sebastian, E. Eleftheriou, *Nat. Nanotechnol.* **2016**, *11*, 693.

[10] T. C. Chong, L. P. Shi, R. Zhao, P. K. Tan, J. M. Li, H. K. Lee, X. S. Miao, A. Y. Du, C. H. Tung, *Appl. Phys. Lett.* **2006**, *88*, 122114.

[11] R. E. Simpson, P. Fons, A. V. Kolobov, T. Fukaya, M. Krbal, T. Yagi, J. Tominaga, *Nat. Nanotechnol.* **2011**, *6*, 501.

[12] X. Zhou, J. K. Behera, S. Lv, L. Wu, Z. Song, R. E. Simpson, *Nano Futur.* **2017**, *1*, 25003.

[13] J. Tominaga, A. V. Kolobov, P. Fons, T. Nakano, S. Murakami, *Adv. Mater. Interfaces* **2014**, *1*, 1300027.

[14]  J. Momand, F. R. L. Lange, R. Wang, J. E. Boschker, M. A. Verheijen, R. Calarco, M. Wuttig, B. J. Kooi, *J. Mater. Res.* **2016**, *31*, 3115.

[15]  J. Kalikka, X. Zhou, E. Dilcher, S. Wall, J. Li, R. E. Simpson, *Nat. Commun.* **2016**, *7*, 11983.

[16]  J. Kim, J. Kim, Y. S. Song, R. Wu, S. H. Jhi, N. Kioussis, *Phys. Rev. B* **2017**, *96*, 235304.

[17]  J. Tominaga, Y. Saito, K. Mitrofanov, N. Inoue, P. Fons, A. V. Kolobov, H. Nakamura, N. Miyata, *Adv. Funct. Mater.* **2017**, *27*, 1702243.

[18]  Y. Saito, K. Makino, P. Fons, A. V. Kolobov, J. Tominaga, *ACS Appl. Mater. Interfaces* **2017**, *9*, 23918.

[19]  A. Caretta, B. Casarin, P. Di Pietro, A. Perucchi, S. Lupi, V. Bragaglia, R. Calarco, F. R. L. Lange, M. Wuttig, F. Parmigiani, M. Malvestuto, *Phys. Rev. B* **2016**, *94*, 045319.

[20]  B. Casarin, A. Caretta, J. Momand, B. J. Kooi, M. A. Verheijen, V. Bragaglia, R. Calarco, M. Chukalina, X. Yu, J. Robertson, F. R. L. Lange, M. Wuttig, A. Redaelli, E. Varesi, F. Parmigiani, M. Malvestuto, *Sci. Rep.* **2016**, *6*, 22353.

[21]  A. Lotnyk, U. Ross, T. Dankwort, I. Hilmi, L. Kienle, B. Rauschenbach, *Acta Mater.* **2017**, *141*, 92.

[22]  S. Cecchi, E. Zallo, J. Momand, R. Wang, B. J. Kooi, M. A. Verheijen, R. Calarco, *APL Mater.* **2017**, *5*, 026107.

[23]  J. Momand, R. Wang, J. E. Boschker, M. A. Verheijen, R. Calarco, B. J. Kooi, *Nanoscale* **2015**, *7*, 19136.

[24]  J. Momand, R. Wang, J. E. Boschker, M. A. Verheijen, R. Calarco, B. J. Kooi, *Nanoscale* **2017**, *9*, 8774.

[25]  A. Yamamoto, T. Mishina, Y. Masumoto, *Phys. Rev. Lett.* **1994**, *73*, 740.

[26]  T. Dekorsy, G. C. Cho, H. Kurz, in *Topics in Applied Physics: Light Scattering in Solids VIII* (Eds: M. Cardona, G.Güntherodt) Springer, Berlin **2000**, pp. 169–204.

[27] R. R. Das, Y. I. Yuzyuk, P. Bhattacharya, V. Gupta, R. S. Katiyar, *Phys. Rev. B* **2004**, *69*, 132302.

[28] M. Nakayama, K. Kubota, T. Kanata, H. Kato, S. Chika, N. Sano, *Jpn. J. Appl. Phys.* **1985**, *24*, 1331.

[29] P. Y. Yu, M. Cardona, in *Fundamentals of Semiconductors*, Springer, Berlin **2005**, pp. 494-502.

[30] C. Colvard, T. A. Gant, M. V. Klein, R. Merlin, R. Fischer, H. Morkoc, A. C. Gossard, *Phys. Rev. B* **1985**, *31*, 2080.

[31] D. J. Lockwood, M. W. C. Dharma-wardana, J.-M. Baribeau, D. C. Houghton, *Phys. Rev. B* **1987**, *35*, 980.

[32] A. Bartels, T. Dekorsy, H. Kurz, K. Kohler, *Appl. Phys. Lett.* **1998**, *72*, 2844.

[33] Y. Saito, P. Fons, A. V. Kolobov, J. Tominaga, *Phys. Status Solidi Basic Res.* **2015**, *252*, 2151.

[34] Y. Saito, P. Fons, L. Bolotov, N. Miyata, A. V. Kolobov, J. Tominaga, *AIP Adv.* **2016**, *6*, 45220.

[35] T. Suzuki, Y. Saito, P. Fons, A. V Kolobov, J. Tominaga, M. Hase, *Appl. Phys. Lett* **2017**, *111*, 112101.

[36] M. Hase, M. Katsuragawa, A. M. Constantinescu, and H. Petek, *Nat. Photon.* **2012**, *6*, 243.

[37] G. Kresse and J. Furthmuller. Phys. Rev. B, 54(16):11169–11186, 1996.

[38] A. Togo and I. Tanaka. Scr. Mater., 108:1 – 5, 2015.

[39] P. E. Blochl. Phys. Rev. B, 50(24):17953–17979, 1994.

[40] G. Kresse and D. Joubert. Phys. Rev. B, 59(3):1758–1775, 1999.

[41] K. Makino, Y. Saito, P. Fons, A. V Kolobov, T. Nakano, J. Tominaga, M. Hase, *Appl. Phys. Lett.* **2014**, *105*, 151902.

[42] Y. Kim, X. Chen, Z. Wang, J. Shi, I. Miotkowski, Y. P. Chen, P. A. Sharma, A. L. Lima Sharma, M. A. Hekmaty, Z. Jiang, D. Smirnov, *Appl. Phys. Lett.* **2012**, *100*, 071907.

[43] D. Brick, V. Engemaier, Y. Guo, M. Grossmann, G. Li, D. Grimm, O. G. Schmidt, M. Schubert, V. E. Gusev, M. Hettich, T. Dekorsy, *Sci. Rep.* **2017**, *7*, 5385.

[44] B. Jusserand, D. Paquet, F. Mollot, F. Alexandre, G. Le Roux, *Phys. Rev. B* **1987**, *35*, 2808.

[45] S. M. Rytov, *Sov. Physics. Acoust.* **1956**, *2*, 68.

[46] A. V. Kolobov, P. Fons, Y. Saito, J. Tominaga, *ACS Omega* **2017**, *2*, 6223.

[47] O. Cojocaru-Miredin, H. Hollermann, M. Wuttig, *Proceeding Eur. Phase Chang. Ovonic Symp.* **2017**, 161.

[48] E. Zallo, R. Wang, V. Bragaglia, R. Calarco, *Appl. Phys. Lett.* **2016**, *108*, 221904.

[49] J. Kalikka, X. Zhou, J. Behera, G. Nannicini, R. E. Simpson, *Nanoscale* **2016**, *8*, 18212.

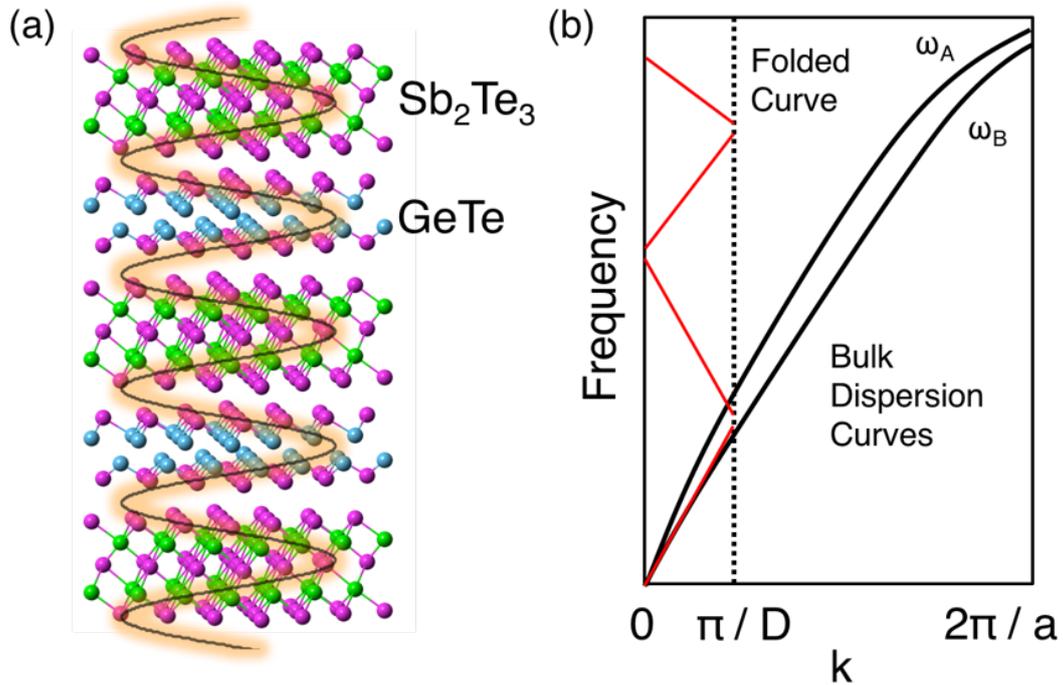

**Figure 1.** (a) Schematics of an FLA phonon mode in an SL structure. The black solid sinusoidal wave represents the FLA phonon, which is not confined into a layer but propagates over the SL structure. The blue balls represent Ge atoms, the purple balls are Te atoms, and green balls are Sb atoms. Note that the Te-Ge-Ge-Te sequence is under debate, but it is just a model structure here. (b) In $k$ space, the FLA phonon dispersion curve is defined as an average of two folded zig-zag lines of the bulk phonon dispersion curve.

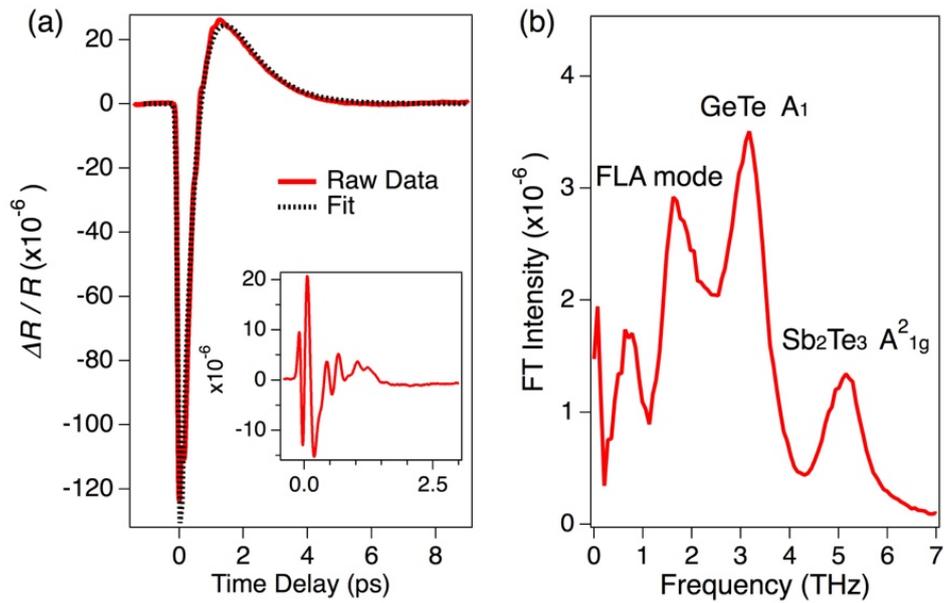

**Figure 2.** (a) Time resolved transient reflectivity change in the sample A. The red curve is a raw data and the black dotted curve is a double-exponential fit. Fitting residual is shown in the inset. (b) Fourier transform of the inset in (a). The FLA, GeTe $A_1$, and Sb$_2$Te$_3$ $A^2_{1g}$ modes are clearly observed.

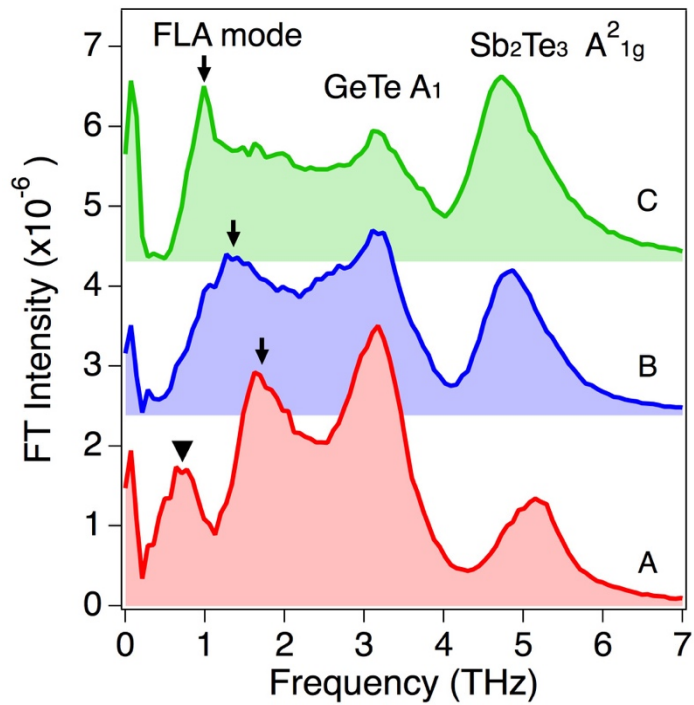

**Figure 3.** Fourier transformed spectra of three samples. The FLA mode frequencies are 1.62, 1.34, 0.99 THz in the sample A, B, and C, respectively, as shown by the black arrows. The inverted triangle represents an additional FLA mode.

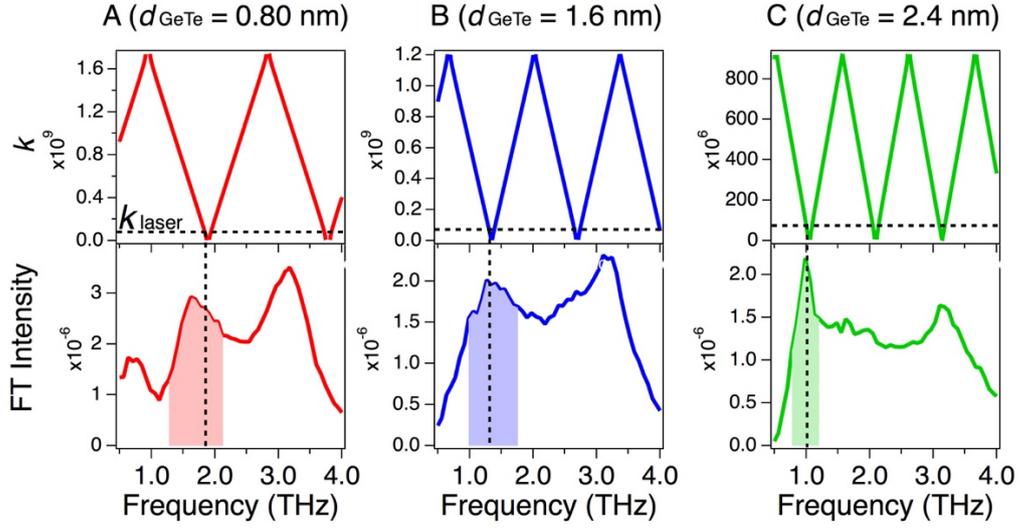

**Figure 4.** (top) Theoretical phonon dispersion curves calculated from Equation (1). The horizontal dotted line is the wavevector of the laser used ($k_{laser}$). (bottom) Experimental FT spectra of three chalcogenide SL samples in low frequency range. The vertical dotted line represents the intersection between $k_{laser}$ and the FLA dispersion. The shaded area in each panel represents the region of interest, which is the FLA mode.

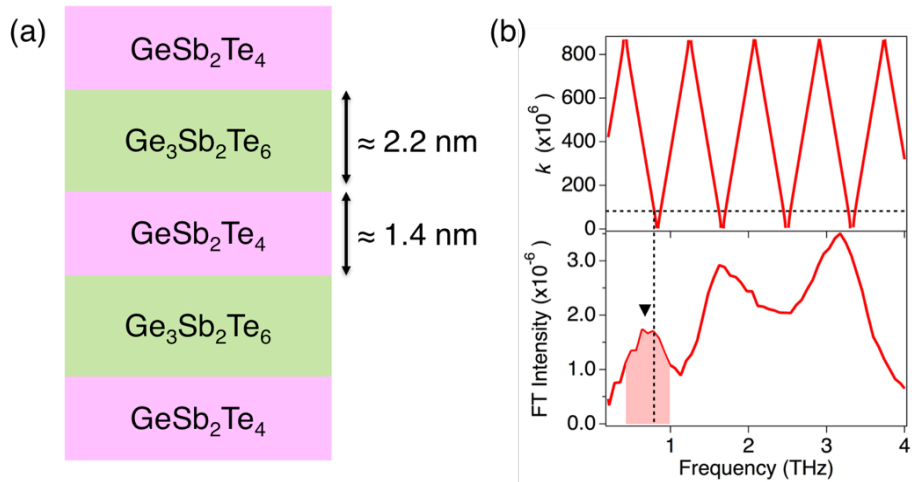

**Figure 5.** (a) A schematic of a possible SL structure, resulting from intermixing in the sample A. (b) (top) The calculated FLA phonon dispersion curve for the structure in (a). The horizontal dashed line is wavevector of the laser. (bottom) The FT spectrum of sample A. A possible FLA phonon peak from the model calculation is indicated by the vertical dashed line, whereas the inverted triangle represents the peak position of the experimental FLA mode. The shaded area represents the region of interest. The second FLA frequency on the dispersion curve seems to match to the peak (~1.6THz), we assign this to FLA mode in $(GeTe)_2/Sb_2Te_3$ SL.